\documentclass{aa} 

\usepackage{longtable}
\usepackage{lipsum} 
\usepackage{graphicx}
\usepackage{txfonts}
\usepackage{multirow}
\usepackage{hyperref}
\usepackage{comment}
\usepackage{microtype}

\makeatletter
\renewcommand*\aa@pageof{, page \thepage{} of \pageref*{LastPage}}

\begin{document}

   \title{XRBcats: Galactic High Mass X-ray Binary Catalogue
   \thanks{The catalogue is available at  CDS via anonymous ftp to cdsarc.u-strasbg.fr (130.79.128.5) or via \url{http://cdsarc.u-strasbg.fr/viz-bin/cat/J/A+A/677/A134}}
   \thanks{A web version is publicly accessible at \url{http://astro.uni-tuebingen.de/\~xrbcat/}}}

   \author{  M. Neumann \thanks{E-mail: marvin.neumann@astro.uni-tuebingen.de}, 
        A. Avakyan,
        V. Doroshenko
        \and 
        A. Santangelo
          }
          
   \institute{Universit{\"a}t T{\"u}bingen, Institut f{\"u}r Astronomie und Astrophysik T{\"u}bingen, Sand 1, T{\"u}bingen, Germany}

   \date{Received 19 December 2022; accepted Accepted 3 July 2023}

 
  \abstract
   {We present a new catalogue of high mass X-ray binaries (HMXBs) in the Galaxy that we call the Galactic High Mass X-ray Binary Catalogue (XRBcats), which improves upon the most recent of such catalogues.
We include new HMXBs discovered since previous publications and revise the classification for several objects previously considered HMXBs or HMXB candidates. The catalogue includes basic information (e.g. source names, coordinates, types),  other data (e.g.  distance and X-ray flux estimates, binary system parameters), and other characteristic properties of the 169 HMXBs catalogued. We also present finding charts in several bands from the infrared to hard X-rays for each object.}
   {The aim of this catalogue is to provide   a list of all currently known Galactic HMXBs, including basic information on both compact objects and non-degenerate counterpart properties (where available). We also include objects tentatively classified as HXMBs in the literature and give a brief motivation for the classification in each case.}
   {The catalogue is compiled based on a search of known HMXBs and candidates in all publicly available databases and literature published before May 2023. The relevant properties in various wavelength bands were collected for all objects, either from the literature or using data provided by large-scale surveys. For the latter case, the counterparts in each individual survey were found by cross-correlating positions of identified HMXBs with relevant databases.}
   {An up-to-date catalogue of Galactic HMXBs is presented to facilitate research in this field. Our goal was to collect a larger set of relevant HMXB properties in a more uniform way compared to previously published works.}
   {}

   \keywords{catalogues -- binaries: close --
                stars: early-type  -- X-rays:
                binaries
               }
   \maketitle
%

\section{Introduction}
X-ray binaries (XRBs) are still the focus of observational X-ray astronomy almost 60 years after their discovery.
These systems consist of a compact object, which can be  a neutron star (NS), a black hole (BH), or a white dwarf (WD), and a non-degenerate companion providing matter for the accretion powering the X-ray emission. Depending on the mass of the optical companion, XRBs can subdivided into two categories. Systems with low mass optical companions ($M_{\mathrm{opt}})\lesssim1M_\odot$ are classified as low mass X-ray binaries (LMXBs \citealt{2022abn..book.....C}),
whereas those with higher mass optical companions ($M_{\mathrm{opt}}\gtrsim 8 M_\odot$) are classified as high mass X-ray binaries (HMXBs, 
\citealt{2022abn..book.....C}).
In the cases where the compact object is either a NS or a BH, accretion is usually assumed. However, systems in which the compact object is a WD are commonly considered separately from the first two cases, and are most often referred to as cataclysmic variables (CVs) rather than XRBs due to both physical and historical reasons. The mass gap between LMXBs and HMXBs is populated by a handful of sources (e.g. Her~X-1 with an optical companion with a  mass of $M_{\mathrm{opt}}\sim2M_\odot$ \citep{1972ApJ...174L.143T}, often referred to as intermediate-mass X-ray binaries (IMXBs, \citealt{2003ApJ...597.1036P}). Despite the mass of the optical companion, all of these binaries host a compact object, and thus represent endpoints of stellar evolution. Understanding the properties of this population is key to understanding the evolution of the Galaxy as a whole. 
Due to their short characteristic lifetimes of $\sim 2\times 10^7 \mathrm{years}$, HMXBs can be considered young systems when compared to LMXBs ($\sim 10^{10} \mathrm{years}$), and are thus good tracers for star forming activity \citep{2003MNRAS.339..793G}. Within the Milky Way such systems are concentrated around the Galactic plane and are often localised within the spiral arms. Similarly to other astrophysical sources, there are still large uncertainties in the spatial distributions, luminosity distributions, numbers, physics, and evolutionary cycles for XRBs. This issue is exaggerated by the fact that we only observe a fraction of the overall population of these systems.
However, the number of known objects is ever increasing. It is therefore essential to keep the census of known XRBs and their properties up to date, especially in the era of new large-scale surveys such as eRosita and \textit{Gaia} that provide a wealth of new observational information across the entire sky. The goal of this paper is therefore to provide such an update for HMXBs. In parallel, similar efforts for LMXBs and IMXBs have been undertaken independently \citep{2023A&A...675A.199A}.

As already mentioned, HMXBs constitute a relatively broad class of objects defined based on the mass of the companion star. Depending on the properties of the companion, and the accretion mechanism, they can be divided into several sub-classes. Systems accreting via stellar winds of their optical companion with mass-loss rates of up to $\Dot{M}\sim 10^{-6..-5} \mathrm{M_\odot \  yr^{-1}}$ \citep{2000ARA&A..38..613K} are usually persistent sources of X-rays and are referred to as supergiant X-ray binaries (SGXBs, \citealt{1986MNRAS.220.1047C}). Within the category of SGXBs are also the highly variable supergiant fast X-ray transients (SFXTs, \citealt{2006ESASP.604..165N}), which are non-accreting most of the time, but show occasional bright flares on timescales of minutes to days with peak fluxes comparable to other SGXBs. Depending on the mass ratio, accretion can be quasi-symmetric, or focused by gravity via the L1 point,  as is the case of Vela~X-1 \citep{2021A&A...652A..95K}. In some cases Roche-lobe overflow (RLO) can also occur; the Galactic micro-quasar SS~433 and the first X-ray pulsar ever discovered, Cen~X-3, are examples where this mechanism dominates. With mass-loss rates of $\Dot{M}\sim 10^{-3} \mathrm{M_\odot \  yr^{-1}}$ (for a fully developed RLO)
these systems have  rather short lifetimes (of the order of the thermal timescale of the optical companion), and are thus rare. However, \citet{1978A&A....62..317S} showed that an atmospheric RLO is feasible and could be sustained for thousands of years with mass-loss rates below $\Dot{M}\sim 10^{-8} \mathrm{M_\odot \  yr^{-1}}$, and indeed more HMXBs of this type are known to exist outside of our own Galaxy. However, the known HMXB population is  dominated by Be X-ray binaries (BeXBs, \citealt{2011Ap&SS.332....1R}). In these systems the optical companion is either an early-type B star (earlier than B3) or a late-type O star  (later than O8), with an equatorial decretion disc. This disc is formed by the material released at the equator of the optical companion due to its high rotation velocity \citep{2020MNRAS.493.2528B}. Decretion disc size is known to be variable and can be traced by the characteristic emission lines in the optical spectra of Be star. Passage of the compact object (in the vast majority of cases a NS) through the disc leads to enhanced accretion, especially if the disc is in an expanding state. From an observational point of view, this leads to the appearance of two different kinds of outbursts that are characteristic of BeXB systems. Type I outbursts are periodically occurring events with a typical peak luminosity below $10^{37} \mathrm{erg\  s^{-1}}$, normally coinciding with the periastron passage of the compact object. Type II outbursts are more irregular, less frequent events, and are apparently associated with the expansion of the circumbinary disc. Peak luminosity can reach up to $10^{39} \mathrm{erg\  s^{-1}}$ \citep{2020MNRAS.491.1857D}, thereby exceeding the Eddington luminosity of the neutron star. Outbursts of this type are usually referred to as giant outbursts and do not show any preferred orbital phase, and usually   last longer than their type I counterparts. Importantly, all BeXBs are transients with relatively low duty cycles and are mostly observable only during their outbursts. They continue, therefore, to be  discovered at a steady rate, and contribute to the increase in the number of known HMXBs.\newline\newline

Prior to the recent catalogue \citet{2023A&A...671A.149F}, the most commonly used catalogue of HMXBs, and the starting point of this work, was \citet{2006A&A...455.1165L}, which was published $\sim$16 years ago. Since then many new HXMBs have been discovered,  and some previously categorised objects have lost their HMXB status. What is perhaps even more important, is that a wealth of additional observational data were collected with new facilities such as the X-ray Multi-Mirror Mission (\textit{XMM-Newton}) \citep{2001A&A...365L...1J}, the International Gamma-Ray Astrophysics Laboratory (\textit{INTEGRAL}) \citep{2003A&A...411L...1W} or \textit{Gaia} \citep{2016A&A...595A...1G}, to name but a few. In this work we present an updated catalogue of Galactic HMXBs including multi-wavelength information as well as the new sources discovered since the publication by \citet{2006A&A...455.1165L}.
The current catalogue only includes Galactic HMXBs and does not include objects from the Small or Large Magellanic Clouds; candidates within these regions often have poorly identified or characterised optical counterparts.
For Galactic HMXBs, a concerted effort has been made within the field to collect all relevant and current multi-wavelength information including distances, optical magnitudes, variability information in soft and hard X-rays, and information on detection in radio band, amongst others. Such information ensures that the catalogue can aid in the identification of new HMXBs in the ongoing and future X-ray surveys, population studies, and other investigations.

\section{Definition of the sample and data sources}\label{sec:Definition}
As a first step, we compiled a sample of 169 HMXBs (123 confirmed and 46 candidates)  by searching the \textit{SIMBAD}\footnote{\url{http://simbad.u-strasbg.fr/simbad/}} and \textit{VizieR}\footnote{\url{https://vizier.cds.unistra.fr/viz-bin/VizieR}} archives hosted by the Centre de Donn\'{e}es astronomiques de Strasbourg (CDS) databases and the literature. The largest fraction of this sample (105 objects) originates from  \citet{2006A&A...455.1165L}, the catalogue this work
is based on. We also systematically searched the literature  (including Astronomoner's Telegrams\footnote{\url{https://astronomerstelegram.org/}}) for reports of new HMXB discoveries. A large fraction of new HMXBs reported in the literature were discovered between 2006 and 2020 by both the \textit{INTEGRAL} mission and the Neil Gehrels Swift Observatory (\textit{Swift}) \citep{2004ApJ...611.1005G}  (29 objects), as summarised in \citet{2019NewAR..8601546K}, which can therefore also be considered  one of the main sources used in this work. As the next step, we considered all objects classified as HXMB or candidates in the \textit{SIMBAD} and \textit{VizieR} databases. Considering that the majority of objects not present in these databases or the literature are extragalactic, and the fact that HMXBs are strongly clustered towards the Galactic plane \citet{2002MmSAI..73.1053G}, we restricted our search to the Galactic latitude between $\pm 15^{\circ}$ in the latter case. 
As an aside, the recent HXMB catalogue \citet{2023A&A...671A.149F} was published during the review process of this paper. We therefore utilised this catalogue as an additional source list, and cross-checked our own catalogue against this, which resulted in an additional five HMXBs.    
We note that although considerable effort was put into finding all possible HXMBs and candidates, it is still possible that some sources have been omitted, we therefore urge the reader to report such omissions to the corresponding author. We also note that some of the objects considered by \citet{2006A&A...455.1165L} as HMXBs or candidates are no longer accepted as such. In Section \ref{sec:Exception} we discuss these cases separately. 

\subsection{HMXB sample and classification}
\begin{figure}[t]   
    \begin{center}
    \includegraphics[width=\columnwidth]{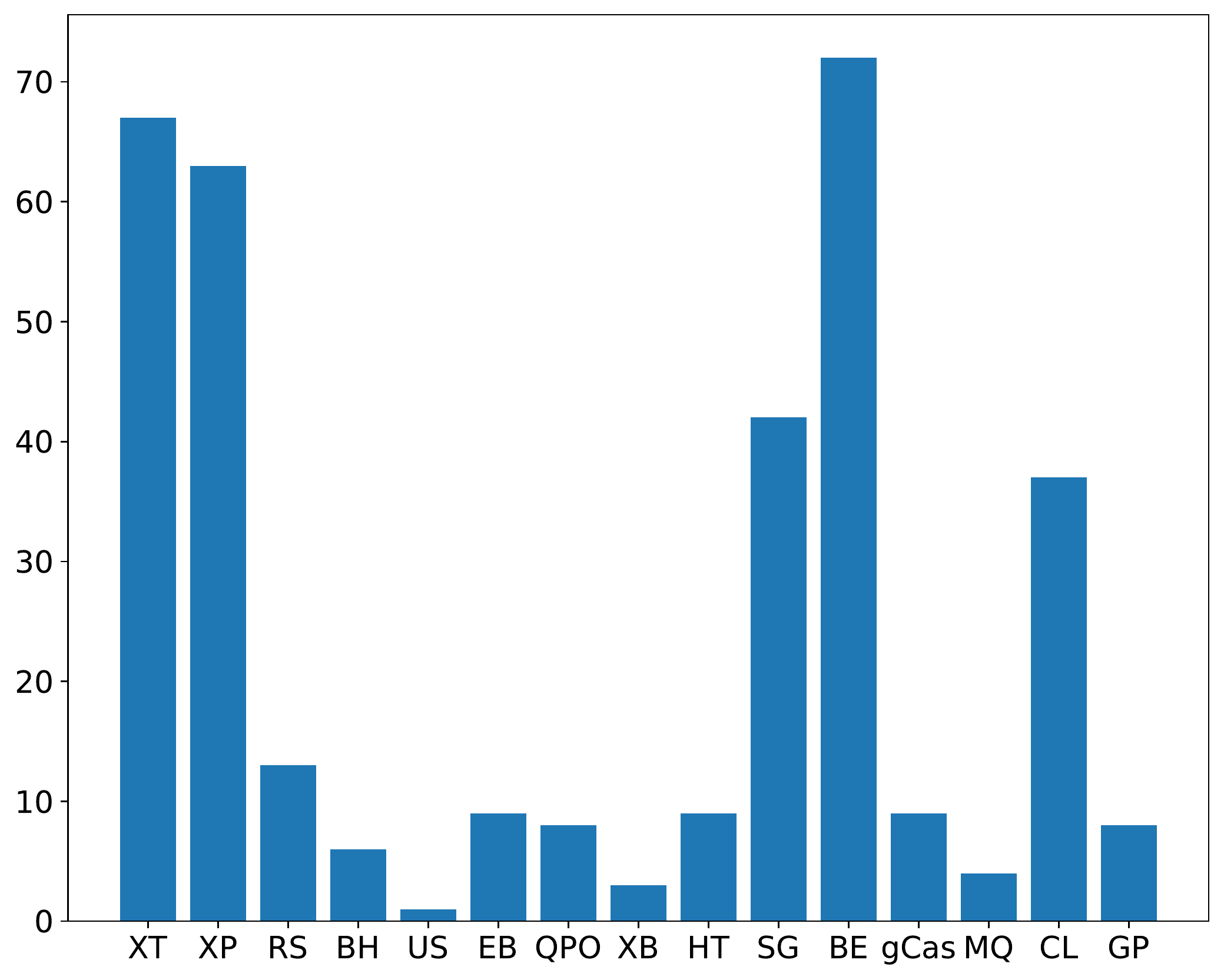}
    \caption{HMXBs population of each type in the Galaxy.} 
    \label{fig:hist}
    \end{center}
    \end{figure}

To reflect the variety among the properties of HMXBs, we include information on the origin of the donor star or compact object, system variability, and other relevant properties. These properties are  then used to define a set of flags designed to represent the range of properties present in this catalogue. In general, the criteria for these flags are based on the classifications introduced in \citet{2006A&A...455.1165L}. However, we have  supplemented them with additional flags based on the information provided within \citet{2019NewAR..8601546K}, \citet{2019A&A...622A..61S}, and \citet{2021MNRAS.507.3899V,2022MNRAS.516.4844V}. As a result, we converged on the set of flags listed below. The frequency of individual flags is provided in brackets and is also presented graphically in Fig.~\ref{fig:hist} to give a broad overview of the sample of known HMXBs. We note  that  the flags below are not meant as a basis for classification of HXMBs, but rather  illustrate the range of different systems within the population considered in this catalogue, meaning a system may have multiple flags:

 \begin{enumerate} \label{em: Flags}
    \item BH: black hole candidate (6);
    \item  EB: eclipsing or partially eclipsing binary system (9);
    \item  MQ: micro-quasar (4);
    \item RS: radio emitting HMXB(13); 
    \item XB: X-ray burst source (3);
    \item XP: X-ray pulsar (63);
    \item  XT: transient X-ray source (67);
    \item  US: ultra-soft X-ray spectrum (1);
    \item HT: hard transient (9);
    \item Gcas: Gamma Cassiopeiae-like source (9);
    \item QPO:  quasi-periodic oscillation (8);
    \item SG: supergiant optical companion (42);
    \item BE: Be star companion (72);
    \item CL: cyclotron resonance scattering feature in X-ray spectrum (37);
    \item GP: high mass gamma-ray binary (8).
 \end{enumerate}

 \subsection{X-ray properties}
 Considering that the prime feature of XRBs is their X-ray emission, we attempted to compile the relevant X-ray properties of the HMXBs in the catalogue. This includes fluxes in the soft X-ray band as observed by \textit{XMM-Newton}, the  Chandra X-Ray Observatory (\textit{Chandra}; \citealt{2000SPIE.4012....2W}),  and the  Swift X-Ray Telescope (\textit{Swift}/XRT; \citealt{2005SSRv..120..165B}), and in the hard   X-ray bands as observed by \textit{INTEGRAL} and the Swift Burst Alert Telescope (\textit{Swift}/BAT; \citealt{2005SSRv..120..143B}).
 We also compiled X-ray positions, in order to assess the reliability of optical counterpart identification or identify plausible counterparts, as described below. In the case of \textit{XMM-Newton} we report flux in the 0.2-12\,keV energy band (corresponding to EPIC\_8 band in the \textit{XMM-Newton} catalogues\footnote{\url{https://heasarc.gsfc.nasa.gov/W3Browse/xmm-newton/xmmssc.html}}  \citealt{2020A&A...641A.136W}).
For \textit{Chandra}\footnote{\url{https://vizier.cds.unistra.fr/viz-bin/VizieR-3?-source=IX/57/csc2master}} \citep{2010ApJS..189...37E} we preferentially used the broad ACIS band ($0.5-7.0$\,keV) or the wide HRC band ($0.1-10.0 \mathrm{keV}$) depending on the availability. 
In the case of \textit{Swift}/XRT we used data that is available in the  Swift X-ray telescope point source catalogue\footnote{\url{https://vizier.cds.unistra.fr/viz-bin/VizieR-3?-source=IX/58/2sxps}} (2SXPS) \citep{2020ApJS..247...54E}, where for the flux we utilised the energy range $0.3-10.0 \mathrm{keV}$, calculated from an assumed power-law spectrum \citep{2020ApJS..247...54E}.
 In the hard X-ray regime we used an energy range of $14-145$\,keV for \textit{Swift}/BAT\footnote{\url{https://heasarc.gsfc.nasa.gov/W3Browse/swift/swbat105m.html}} \citep{2018ApJS..235....4O} data and an energy range of $17-60$\,keV for \textit{INTEGRAL} data\footnote{\url{https://vizier.cds.unistra.fr/viz-bin/VizieR-3?-source=J/A\%2bA/545/A27}} \citep{2012A&A...545A..27K}. 

\subsection{Optical position and astrometry}
 Most of the HMXBs in the sample have identified optical counterparts (an almost essential pre-requisite for HXMB classification). However, the optical positions reported in \citet{2006A&A...455.1165L} are often limited by the techniques of the time, and thus show discrepancies with modern observations. An effort was made to update these data with the updated astrometry provided by the \textit{Gaia} mission. To achieve this, we first confirmed whether the \textit{SIMBAD} database already contained identified \textit{Gaia}~DR3\footnote{\url{https://vizier.cds.unistra.fr/viz-bin/VizieR-3?-source=I/355}} \citep{2023A&A...674A...1G} or Two Micron All Sky Survey\footnote{\url{https://vizier.cds.unistra.fr/viz-bin/VizieR?-source=II/246}} (2MASS; \citealt{2003tmc..book.....C}) counterparts. If this was not the case, we matched the literature positions, which included V-band magnitudes, to \textit{Gaia}~DR3 using a large search radius of 10$^{\prime\prime}$. 
 \textit{Gaia} counterparts that matched the literature positions and magnitudes within the uncertainties were then selected.
 In most cases this corresponded to an object closest to the literature position with  V-band magnitude within 2\,mag of the G-magnitude reported by \textit{Gaia}. We note that most of the HMXBs are, to some extent, variable in the optical band so an exact comparison would be unfeasible. This meant that the threshold of two~magnitudes was determined empirically by inspecting the observed difference between the  literature and \textit{Gaia} magnitudes for sources where the \textit{Gaia} counterpart was already unambiguously identified.
 
For all HMXBs with a \textit{Gaia} counterpart, we provide distance information from \citet{2023A&A...674A...1G}, \citet{2021AJ....161..147B},\footnote{\url{https://vizier.cds.unistra.fr/viz-bin/VizieR-3?-source=I/352}} and \citet{2022A&A...658A..91A}.\footnote{\url{https://vizier.cds.unistra.fr/viz-bin/VizieR-3?-source=I/354/starhorse2021}} In addition, for objects without an identified \textit{Gaia} counterpart, but with available distance estimates in the literature, we  give those estimates and their corresponding references. If multiple distance estimates were available, the mean distance was calculated by using the arithmetic mean on all available distance estimations linked to the given source. The range of possible distances, calculated by taking the lowest and highest estimations of all distance estimates for a given system (from the literature or from \textit{Gaia}) is also reported.

 \subsection{Additional information}
 We also attempted to include some extra information related to the properties of the optical counterparts and the compact objects.  
 For instance, in addition to the G-band magnitudes of \textit{Gaia} published in \citet{2023A&A...674A...1G}, magnitudes in the J-, H-, and K-bands provided by \citet{2003tmc..book.....C} as well as magnitudes in the W1- and W2-bands of the Wide-field Infrared Survey Explorer mission \citep{2010AJ....140.1868W} in CatWISE2020\footnote{\url{https://vizier.cds.unistra.fr/viz-bin/VizieR?-source=II/365}} \citep{2021ApJS..253....8M} and available V-band magnitudes were included. 
 Any luminosity estimates of the optical companions from \citet{2023A&A...674A...1G} were also added. 
 Since the spectral type of the counterpart is also pertinent information, we attempted to include this information by using the following sources: \citet{2019NewAR..8601546K}, \citet{2019A&A...622A..61S}, \citet{2006A&A...455.1165L}, Mauro Orlandini's website,\footnote{\url{http://www.iasfbo.inaf.it/~mauro/pulsar_list.html}}
 and \textit{SIMBAD}. 
 Due to the close connection between spectral type and effective stellar temperature, estimations of the temperatures provided by \citet{2022A&A...658A..91A} and \citet{2023A&A...674A...1G} were included. The catalogue includes only the mean stellar effective temperature.
 One of the key properties of NS systems is the magnetic field strength of the compact object. This can be measured using observed energies of the cyclotron resonance scattering features (CRSFs, or cyclotron lines). We therefore included the literature values of the observed CRSF energies where such features were claimed to be detected. This includes the fundamental line as well as harmonics. The line energies and corresponding references were sourced from the recent review by \citet{2019A&A...622A..61S} and the X-ray pulsar properties database by Mauro Orlandini, which appears to cover all X-ray pulsars where a detection of a CRSF was claimed in the literature. 

\section{Catalogue content and quality assurance}
\subsection{Description of the fields}
The Galactic High Mass X-ray Binary catalogue (XRBcats) we present here  
contains a total of 169 HMXBs and candidates, sorted according to increasing right ascension (second column).
The table consists of 66 columns, listing the various parameters and references for a given HMXB. The first column contains the source name, as decided by the most common name in the literature (number of mentions in NASA ADS system). The following seven columns are dedicated to the coordinates of the system, the second and third columns displaying the right ascension (RA) and declination (DEC) in degrees, followed by statistical uncertainties of the coordinates in arcseconds. The Coord\_Ref column provides the reference of the catalogue or paper, which was used to extract the coordinates and uncertainties (e.g. \citet{2023A&A...674A...1G} for \textit{Gaia} DR3), given as a NASA ADS bibcode. The sixth column indicates whether the optical counterpart is solidly identified in the literature (0), if it is a tentative optical counterpart identified as such in the literature (1) or if it is not yet identified (2). In the seventh and eighth columns the galactic longitude (GLON) and galactic latitude (GLAT) in degrees can be found. The following column displays the different X-ray flags related to HMXB classification, which were discussed in Sect. \ref{sec:Definition}. Columns 10 and 11 provide the orbital period of the compact object in days and the spin period of the pulsar (where appropriate) in seconds, respectively. For NS HMXBs with CRSFs reported in the literature, line energies are listed in column 12 in units of keV regardless of if it is reporting on the fundamental line or any harmonics. In the Alt\_name column, the second most-used identifier in the literature was selected as an alternative name for the HMXB. Column 14 displays the spectral type of the optical companion. If it was possible to identify the \textit{Gaia} counterpart, the \textit{Gaia}-DR3-ID is listed in column 15. Column 16 and 17 show the magnitude in the G-band as well as its uncertainty followed by the V-band magnitude in Column 18 and its uncertainty in Column 19. In Columns 20-29, the JHK-magnitudes as well as the W1- and W2-magnitudes with their respective uncertainties are shown.
The neutral hydrogen column density in units of $10^{21} \mathrm{cm^{-2}}$ is shown in column 30, which is available in \citet{2020ApJS..247...54E}. In this reference hardness ratios were first taken and utilised to produce a power-law spectrum through interpolation; from this interpolated spectrum, the neutral hydrogen column density best fitting that spectrum was derived. 
The following 11 columns contain information on the X-ray flux. Columns 31 and 32 are the minimum and maximum flux values of the source in the XMM-Newton catalogue.
Columns 33 and 34 are the flux values of the source in Chandra, due to the use of both Chandra HRC and ACIS; column 35 indicates which of these two instruments was used for the flux values.
Columns 36 and 37 respectively display the minimum and maximum flux values of \textit{Swift}/XRT of the source, columns  38 and 39 for \textit{Swift}/BAT, and Columns  40 and  41 for INTEGRAL.
In cases where only one observation was taken the same value was adopted for minimum and maximum flux values.
Columns 42-44 contain mass estimates of the compact object,  in particular mean mass, as well as the upper and lower mass limits reported in the literature (in $\mathrm{M}_\odot$).
Columns 45-47 contain distance estimates (in parsec) to a given source, including the mean distance and the upper and lower limits given in the literature. The stellar effective temperature and the luminosity of the optical counterpart can be found in columns 48 and 49, where  $T_{eff}$ is in Kelvin and the luminosity of the companion is in $L_\odot$. All \textit{SIMBAD} identifiers associated with HMXBs are quoted in column 50. The next seven columns are dedicated to the references of CRSF, pulsation period, orbital parameter, spectral type, distance-estimation (if literature values were used), mass estimation, and miscellaneous references. Column 58 is dedicated to comments (e.g. if the source is considered an HMXB candidate or similar). The final eight columns contain the source name given in the following catalogues:  2MASS All-Sky Catalogue of Point Sources,  CatWISE2020 catalogue, Second \textit{ROSAT} all-sky survey (2RXS) source catalogue, \textit{XMM-Newton} Serendipitous Source Catalogue, Chandra Source Catalogue (CSC) Release 2.0, 2SXPS Swift X-ray telescope point source catalogue, \textit{Swift}/BAT  105-Month All-Sky Hard X-Ray Survey catalogue, and \textit{INTEGRAL/IBIS} 9-year Galactic Hard X-Ray Survey
catalogue.

\subsection{Finding charts and problematic cases}\label{sec:Exception}
\begin{figure*}[t]   
    \begin{center}
    \includegraphics[width=\textwidth]{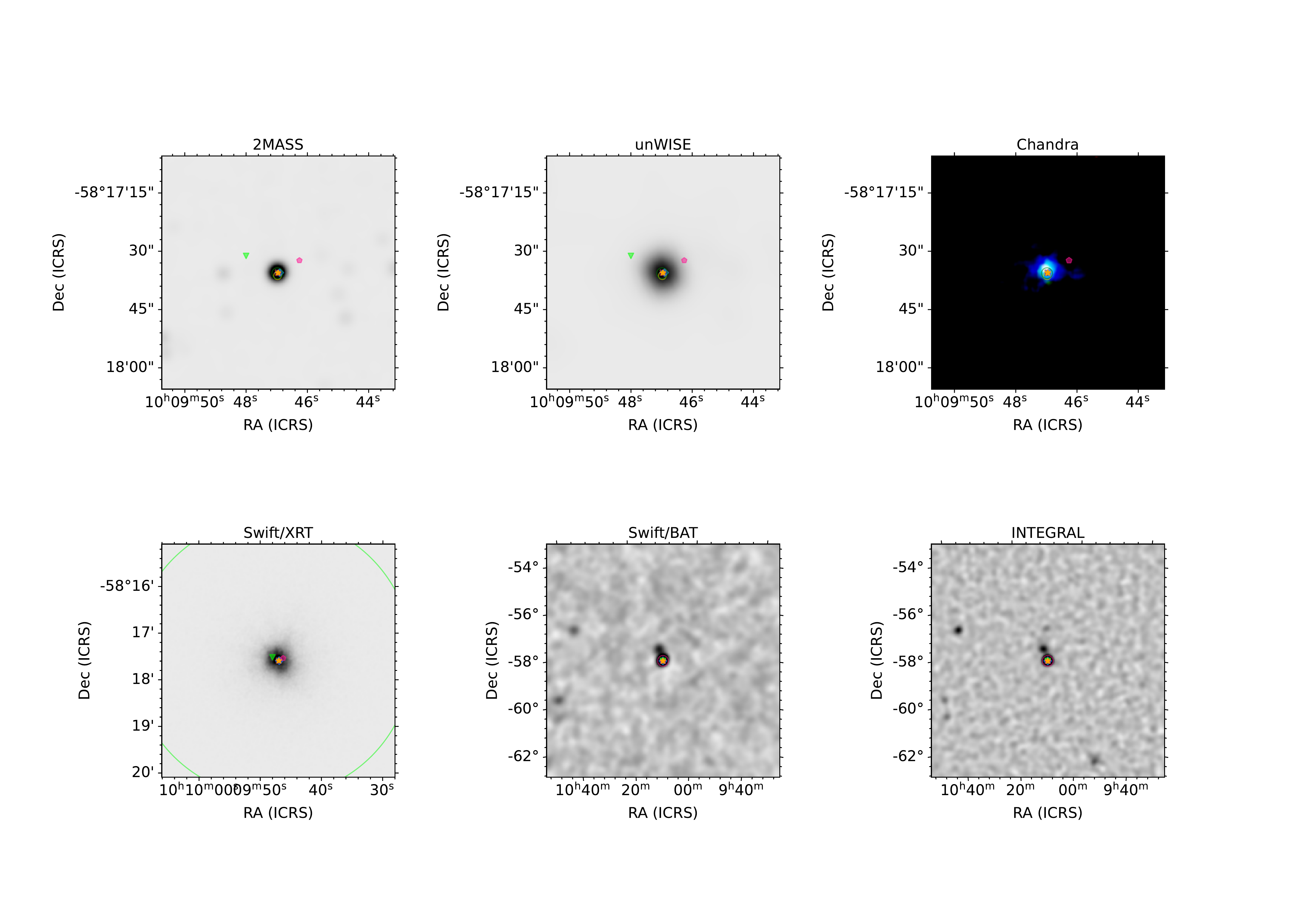}
    \caption{Finding charts of GRO~J1008$-$57. The finding charts are overlaid with symbols and error circles to indicate observations of different instruments. See text in Sect. \ref{sec:Exception} for an explanation of the finding charts and the symbols.
    } 
    \label{fig:finding}
    \end{center}
    \end{figure*}

As part of the catalogue we provide finding charts, which consist of up to six different images ranging from near-infrared (NIR) to hard X-rays for all sources. For the majority of the finding charts, we used the Hierarchical Progressive Surveys (HiPS, \citealt{2015A&A...578A.114F}); only in the case of \textit{Swift}/XRT did we use the SkyView Query. Both HiPS and SkyView offer the possibility to query data automatically with Python. To access HiPS,  and SkyView we used the astroquery.hips2fits\footnote{\url{https://alasky.u-strasbg.fr/hips-image-services/hips2fits}} package and the  astroquery.skyview\footnote{\url{https://astroquery.readthedocs.io/en/latest/skyview/skyview.html}} package, respectively. In the following we give details of the surveys used and provide their corresponding links as footnotes. The finding charts consist of up to six different surveys, which are mentioned below, along with their corresponding links as footnotes. 
In the top row we include the image taken from the Visible and Infrared Survey Telescope for Astronomy (VISTA) from their  Variables in the the Via Lactea (VVV\footnote{\url{http://alasky.cds.unistra.fr/VISTA/VVV_DR4/VISTA-VVV-DR4-J/}}) DR4 catalogue \citep{2010NewA...15..433M}. Alternatively, for the top row image, where VVV was not available we utilised 2MASS.\footnote{\url{http://alasky.cds.unistra.fr/2MASS/J/}} The central image is taken from \textit{unWISE}\footnote{\url{http://alasky.cds.unistra.fr/unWISE/W1/}} \citep{2019ApJS..240...30S} and for the right-hand RGB image we utilised   \textit{Chandra},\footnote{\url{https://cdaftp.cfa.harvard.edu/cxc-hips/}} \textit{XMM-Newton},\footnote{\url{http://skies.esac.esa.int/XMM-Newton/EPIC-RGB/}} or the Roentgensatellit (\textit{ROSAT};\footnote{\url{http://alasky.cds.unistra.fr/RASS/}} \citealt{1982AdSpR...2d.241T, ROSAT, ROSAT99}) (listed here in order of preference). In the left-hand corner of the  bottom row we provide a soft X-ray image of \textit{Swift}/XRT (SwiftXRTInt in astroquery.skyview). For hard X-rays we take images from \textit{Swift}/BAT\footnote{\url{http://cade.irap.omp.eu/documents/Ancillary/4Aladin/BAT_14_20/}} (middle section of bottom row); and \textit{INTEGRAL}\footnote{\url{http://cade.irap.omp.eu/documents/Ancillary/4Aladin/INTEGRAL_17_60/}} (right hand section of bottom row). A 1 arcmin field of view was used for the creation of the VVV, 2MASS, \textit{unWISE}, \textit{XMM-Newton}, and \textit{Chandra} images. In the case of \textit{Swift}/XRT and \textit{ROSAT}, we used a field of view size of 5 arcmin and 15 arcmin, respectively. The field of view for \textit{Swift}/BAT and \textit{INTEGRAL}  was chosen to be $10^\circ$. In each case, the size of the region was chosen considering the field of view and angular resolution of a given instrument. Every panel also shows the coordinates and uncertainties of all detected sources within their respective regions (position uncertainties are represented by error circles).
A red cross indicates the position given by \textit{SIMBAD} for the source, and a dodger-blue diamond for the coordinates described in the literature. The position of the source used by this catalogue is indicated with an orange star. In the case of the soft X-ray instruments, the position of the observations is indicated by the error circles to prevent overcrowding, a golden circle indicates the \textit{Chandra} position, a red circle \textit{XMM-Newton}, a green circle \textit{Swift}/XRT, and a navy blue circle \textit{ROSAT}. \textit{Swift}/BAT and \textit{INTEGRAL} are denoted by a deep pink pentagon and a lime green triangle, respectively. In the optical band we indicate \textit{CatWISE} with an orange X, 2MASS with a cyan plus sign, and \textit{Gaia}~DR3 data with purple square.
As an example, Figure \ref{fig:finding} shows the finding chart of GRO~J1008$-$57.
    
Based on visual inspection of the finding charts and literature research, several problematic cases were identified. We list these cases in their respective categories, in ascending order of their right ascension coordinates. 
Several sources initially assumed to be HMXBs were removed from the sample.
\paragraph{1E~1048.1-5937:}
1E~1048.1-5937 is listed as an HMXB in \textit{SIMBAD}; however, \citet{2009ApJ...702..614D} concluded that the source is actually a magnetar. The source is also listed as a magnetar in the McGill magnetar database\footnote{\url{https://www.physics.mcgill.ca/~pulsar/magnetar/main.html}} \citet{2014ApJS..212....6O}. 
\paragraph{IGR~J18151-1052:}
IGR~J18151-1052 was suggested to be an HMXB \citep{2009ATel.2193....1B}; however, this was shown not to be the case by \citet{2012AstL...38....1L}. The object might be associated with the magnetar candidate PSR~J1845-0258 or be a CV, but in any case its identification as an HMXB is unlikely, so we decided not to include this source.
\paragraph{1E~2259+58.6:}
    Another magnetar 1E~2259+58.6 \citep{2002ApJ...567.1067G} was excluded using the same logic as for 1E~1048.1-5937.\\
 
Several objects  have tentative or known optical counterparts in the literature, but they could not be found automatically through the procedure outlined above. For these objects, the literature was searched manually and the known counterparts and their coordinates were added to the catalogue.

    \paragraph{RX~J0148.9+6121:} Not much is known about RX~J0148.9+6121 which was only referred to once by \citet{2014RAA....14..673W} as having HMXB status. We therefore consider it to be a candidate rather than a confirmed HMXB.
    \paragraph{PSR~J0635+0533:}\citet{2022A&A...665A..31F} compiled a list of  galactic HMXBs that contain a NS as their compact object, and made an effort to identify optical counterparts for the systems in Gaia DR3. The  study only lists objects with unambiguously identified counterparts both in \textit{Gaia} DR3 and 2MASS catalogues that are within $0.5"$ of each other. In the case of PSR~J0635+0533, they identified \textit{Gaia}~DR3~3131755947406031104 as the counterpart.
    \paragraph{1H~0749-600:} For 1H~0749-600, \citet{2001A&A...377..148T} posited that the proposed optical counterpart HD~65663 is not the real counterpart, but rather a single Be star spatially coincident with an X-ray transient. \citet{2001A&A...377..148T} also considered the possibility that the system is not an X-ray binary at all. We therefore classify this association as tentative.
    \paragraph{1FGL~J1018.6-5856:} For 1FGL~J1018.6-5856 the same procedure was performed as in the case of PSR~J0635+0533. Here \citet{2022A&A...665A..31F} identified \textit{Gaia}~DR3~5255509901121774976 as the \textit{Gaia} counterpart for 1FGL~J1018.6-5856.
    \paragraph{1ES~1210-64.6:} Based on the coordinates of the soft X-ray counterpart of 1ES~1210-64.6 \citep{2007ATel.1253....1R}, \citet{2009A&A...495..121Mx} identified the optical counterpart (\textit{Gaia}~DR3~6053076566300433920) This conclusion was confirmed by follow-up optical spectroscopy.
    \paragraph{IGR~J12341-6143:} A soft X-ray counterpart for IGR~J12341-6143 was discovered quite recently by \citet{2020ATel14039....1S} with \textit{Swift}/XRT, which also led to the identification of a tentative optical counterpart (\textit{Gaia}~DR2~6054778507172454912).
    \paragraph{1A~1238-59:} We suggest  classifying 1A~1238-59 as an HMXB candidate, due to an error radius of $30"$ \citep{1978Natur.273..364D} and because there is  no detected optical counterpart.
    \paragraph{IGR~J14059-6116:} We decided to classify IGR~J14059-6116 as  an HMXB candidate, due to a missing optical counterpart. \citet{2019ApJ...884...93C} suggested that IGR~J14059-6116 has a possible association with the HMXB 4FGL~J1405.1-6119, and therefore we  note that IGR~J14059-6116 is likely to be the same source as 4FGL~J1405.1-6119.
    \paragraph{MAXI~J1409-619:} \citet{2010ATel.2962....1K} located an uncatalogued X-ray source inside the MAXI error circle of MAXI~J1409-619 using \textit{Swift}/XRT. Inside the \textit{Swift}/XRT error circle of this source, a catalogued IR source (2MASS~J14080271-6159020) was also found which is considered   a tentative optical counterpart.
    \paragraph{IGR~J14331-6112:} \textit{Gaia}~DR3~5878377736381364608 is mentioned as the unambiguous optical counterpart of IGR~J14331-6112 by \citet{2022A&A...665A..31F}.
    \paragraph{Cir~X$-$1:} Since its discovery, Cir~X$-$1 has been referred to as an LMXB, until \citet{2013ApJ...779..171H}  determined the age of the system to be about $4500$ yr. In addition, according to \citet{2007MNRAS.374..999J} this system contains an A0- to B5-type supergiant companion. However, Cir~X-1 has also shown type I X-ray bursts~\citep{1986MNRAS.221P..27T}, which indicates that source is an LMXB. The possible LMXB origin   is also supported by the fact that the companion star itself in Cir~X-1 still cannot be unambiguously detected at optical wavelengths.
    \citet{2016MNRAS.456..347J} concluded that the donor could either be  a low mass star that has not had time to evolve or  a giant star still recovering from the impact of a supernova blast that happened less than 5000 yr ago. Taking into account all this obscurity around Cir~X-1's nature we decided to add it to both LMXB and HMXB catalogues. 
    \paragraph{XTE~J1543-568:} In March 2012 \textit{Swift}/BAT detected an increased flux from XTE~J1543-568, which led to a more accurate determination of the source position. This allowed \citet{2012ATel.4008....1K} to identify a tentative counterpart (2MASS~J15440515-5645425) within the error box provided by \textit{Swift}/XRT.
    \paragraph{2S~1553-542:} With the help of \textit{Swift}/XRT and \textit{Chandra} observations, \citet{2016MNRAS.462.3823L} were capable of identifying a likely infrared counterpart for 2S~1553-542 at  RA(J2000)=15h 57m 48.28s DEC(J2000)=-54$^\circ$24' 53.5".
    \paragraph{IGR~J16374-5043:} Inside the error circle around the \textit{Swift}/XRT position of IGR~J16374-5043, \citet{2020MNRAS.491.4543S} found only one infrared source. This source is not listed in the 2MASS catalogue but is in the catalogue of \textit{Gaia} (\textit{Gaia}~DR3~5940285090075838848) and is likely to be the optical counterpart of IGR~J16374-5043.
    \paragraph{XTE~J1716-389:}  \citet{2010MNRAS.408.1866R} used \textit{Chandra} for localisation of Galactic X-ray sources and did follow-up observations in the optical and NIR bands. One of the observed sources was XTE~J1716-389. The refined \textit{Chandra} position allowed \citet{2010MNRAS.408.1866R} to identify the only possible optical counterpart, an infrared source 2MASS~J17155645-3851537, that is thus considered to be XTE~J1716-389's counterpart.
    \paragraph{IGR~J17375-3022:} IGR~J17375-3022 was a poorly studied source until \citet{2020MNRAS.491.4543S} investigated it in the hard and soft X-ray bands, and in the infrared band. With an enhanced source position from \textit{Swift}/XRT, the study found only one NIR object within the \textit{Swift}/XRT error circle in the VVV survey. Therefore, \citet{2020MNRAS.491.4543S} proposed that VVV~J173733.74-302314.5 is the best counterpart candidate for IGR~J17375-3022.
    \paragraph{RX~J1739.4-2942:} RX~J1739.4-2942, originally identified as an LMXB, could potentially be a Be/HMXB \citet{2016ATel.8704....1B}. Therefore, we included RX~J1739.4-2942 in the catalogue as an HMXB candidate  (Number 67 in \citealt{2006A&A...455.1165L}).
    \paragraph{IGR~J18029-2016:} \citet{2006A&A...453..133W} proposed two possible NIR counterparts for  IGR~J18029-2016; they were 2MASS~J18024194-2017172 and 2MASS~J180242.0-201720.2, which were found in the 2MASS catalogue and the Second Guide Star Catalogue, respectively. The first candidate, 2MASS~J18024194-2017172, was later confirmed to be the counterpart of the HMXB by \citet{2008A&A...482..113M}.
    
    \paragraph{IGR~J18179-1621:} With the refined position of IGR~J18179-1621 reported by \citet{2012MNRAS.426L..16L}, \citet{2012ApJ...757..143N} were able to observe the object with \textit{Chandra} and identify a possible optical counterpart. 2MASS~J18175218-1621316 was identified as a possible companion, due to it being the only NIR source in the 2MASS catalogue within a $1"$ radius of the \textit{Chandra} position.
    \paragraph{IGR~J18219-1347:} Quite recently \citet{2022ApJ...927..139O} identified a bright IR counterpart for IGR~J18219-1347 close to the \textit{Chandra} localisation, at RA(J2000)=18h 21m 54.821s DEC(J2000)=-13$^\circ$47' 26.703", which appeared to be the combination of two point sources (Star A and Star B). They concluded that Star A is the Counterpart of IGR~J18219-1347, because it is consistent with the SED of a Be star, and therefore the system was classified as a BeXRB.
    \paragraph{AX~J1841.0-0536:} \citet{2022A&A...665A..69F} queried the confirmed HMXBs of their previous work \citep{2022A&A...665A..31F} in \textit{Gaia} DR3. For AX~J1841.0-0536 the source \textit{Gaia}~DR3~4256500538116700160 was therefore identified as an optical counterpart.
    \paragraph{\textit{Ginga}~1839-04:} For \textit{Ginga}~1839-04 no optical counterpart has  yet been identified. \citet{2006A&A...455.1165L} included this source in their catalogue due to a tentative pulsation detection of $\sim 81 \mathrm{s}$ \citep{1990Natur.343..148K} by \textit{Ginga} during an outburst in 1989. Since then no further X-ray detections have been reported \citep{2009ApJ...697.1194S}. Therefore, \textit{Ginga}~1839-04 is classified as a candidate HMXB in this catalogue.
    \paragraph{IGR~J18482+0049:} \citet{2012ApJ...753....3B} observed five \textit{INTEGRAL} sources towards the Scutum Arm, one of which  was IGR~J18482+0049. With the refined position, they found only one object in the 2MASS catalogue that was consistent with the \textit{XMM-Newton} position. Based on this observation, IGR~J18482+0049 has a possible association with 2MASS~J18481540+0047332.
    \paragraph{\textit{Ginga}~1855-02:} Within \textit{Ginga}~1855-02, the positional error is $10'$ \citep{1990Natur.343..148K}. Like \textit{Ginga}~1839-04, \textit{Ginga}~1855-02 is a poorly studied source that does not have an identified optical counterpart, and therefore it is also classified as a candidate HMXB based on its transient behaviour.
    \paragraph{XTE~J1859+083:} After an outburst of XTE~J1859+083 detected by MAXI \citep{2015ATel.7034....1N}, follow-up \textit{Swift}/XRT observations by \citet{2015ATel.7067....1L} allowed the improvement of the position of the HMXB. No UVOT counterpart was detected, but with the enhanced position \citet{2015ATel.7067....1L} were able to find a possible counterpart in the USNO-B1.0 (USNO-B1.0~0982-0467424) and 2MASS catalogues (2MASS~J18590163+0814444).

    \paragraph{1E~1912.5+1031:} \citet{2011ATel.3326....1B} identified 2MASS~J19145680+1036387 as the most likely counterpart of 1E~1912.5+1031, based on the position inside  the error circle of their estimated \textit{Swift}/XRT position for 1E~1912.5+1031.
    
    \paragraph{AX~J1949.8+2534:} For AX~J1949.8+2534 the source 2MASS~J19495543+2533599 was proposed to be the optical counterpart \citep{2017MNRAS.469.3901S} . This was later confirmed to be the case by \citet{2019ApJ...878...15H}. 
    \paragraph{W63~X-1:} W63~X-1 is an X-ray binary within the supernova remnant W63, with a pulsation period of 36 sec \citep{2004HEAD....8.1730R}. Based on the pulsation period and spectrum, \citet{2004HEAD....8.1730R} concluded that the companion is likely an isolated neutron star, HMXB, or LMXB.
    They found an optical counterpart with $H\alpha$-excess emission typical for a Be companion, and thus classified the source as an HMXB.\\

Several objects present in \citet{2006A&A...455.1165L} were removed from the current catalogue as they are no longer classified as HMXBs:

\paragraph{1WGA~J0648.0-4419:} 1WGA~J0648.0-4419, previously listed as source 18 in \citet{2006A&A...455.1165L}, was removed because the optical star in the system has since been classified as a hot sub-dwarf \citep{1963PASP...75..365J}.
\paragraph{IGR~J12349-6434:} IGR~J12349-6434 (source 35 in \citet{2006A&A...455.1165L}] was initially classified as a new source. However, it is thought to be associated with RT~Cru \citep{2005ATel..591....1T}, a symbiotic star containing a WD \citep{2007ApJ...671..741L}. It was therefore excluded from the catalogue.
\paragraph{1A~1246-588:} 1A~1246-588, which was source 38 in \citet{2006A&A...455.1165L}, is an LMXB \citet{2006A&A...446L..17B}.
\paragraph{SAX~J1452.8-5949:} With the de-reddened magnitudes of their observations in the JHK-bands and the assumption of a black-body model, \citet{2009MNRAS.394.1597K} made estimations of the distance of SAX~J1452.8-5949 (source 46 in \citealt{2006A&A...455.1165L}). This ruled out the possibility of the object being an HMXB, due to the fact that the system would be an extragalactic source. They concluded that the binary system must have a low mass companion, and therefore is either an LMXB or an intermediate polar (IP,  accreting magnetised white dwarf).
\paragraph{IGR~J16358-4726:} \citet{2008A&A...484..783C} classified IGR~J16358-4726 (source  55 in \citealt{2006A&A...455.1165L}) as an HMXB; however, this classification was revoked by \citet{2010A&A...516A..94N} and the source is now considered to be a symbiotic X-ray binary.
\paragraph{AX~J1700.1-4157:} Based on NIR observations, \citet{2010MNRAS.402.2388K} estimated that the companions of AX~J1700.1-4157\footnotemark[26] (source 63 in \citealt{2006A&A...455.1165L}) should be a low mass star. In combination with a detected Fe emission line, \citet{2010MNRAS.402.2388K} concluded that AX~J1700.1-4157 is most likely an IP, and  therefore AX~J1700.1-4157 was removed from this catalogue.
\paragraph{IGR~J17091-3624:} IGR~J17091-3624  (source 64 in \citealt{2006A&A...455.1165L}) is also now classified as an LMXB \citep{2016ATel.8761....1G}.
\paragraph{AX~J1740.1-2847:} AX~J1740.1-2847 (source 68 in \citealt{2006A&A...455.1165L}) is a similar case to AX~J1700.1-4157. \citet{2010MNRAS.402.2388K} also estimated that the companion should be a low mass star, and due to a detected Fe emission line they concluded that AX~J1740.1-2847 is most likely an IP as well, and hence it was removed from the catalogue. 
\paragraph{4U~1807-10:} 4U~1807-10 (source 75 in \citealt{2006A&A...455.1165L}) was detected by \textit{UHURU} \citep{1978ApJS...38..357F} within a large error of 1.3$^\circ$.
    To date, it is not certain if 4U~1807-10 is an HMXB or an LMXB. However, it is more likely that 4U~1807-10 is an LMXB as the system shows type I X-ray bursts \citep{2017AstL...43..781C}. Therefore, we   excluded it from this catalogue.
        
\paragraph{SAX~J1819.3-2525:} \citet{2014ApJ...784....2M} determined a mass of 2.9 $M_\odot$ for the optical companion of SAX~J1819.3-2525\footnotemark[26] (source 77 in \citealt{2006A&A...455.1165L}), which classifies this system as an IMXB, and therefore it was excluded from the catalogue.
\paragraph{AX~J1838.0-0655:} AX~J1838.0-0655\footnotemark[26] (source 82 in \citealt{2006A&A...455.1165L}) is currently classified as a supernova remnant. \citet{2005ApJ...630L.157M} does not exclude the possibility of AX~J1838.0-0655 being an X-ray binary, but mentions that this scenario is unlikely due to the lack of strong X-ray and $\gamma$-ray variability, as well as the extension of the source to up to TeV energies. Therefore, AX~J1838.0-0655 was removed from the catalogue.
\paragraph{XTE~J1901+014:} \citet{2008AstL...34..753K} proposed that XTE~J1901+014 (source 94 in \citealt{2006A&A...455.1165L}) could be the first low mass fast X-ray transient. To date, the nature of this system is not completely clear \citep{2019ATel13328....1S}; we therefore decided to exclude this source from the current version of this catalogue.
\paragraph{XTE~J1906+09:} XTE~J1906+09 (source 96 in \citealt{2006A&A...455.1165L}) was not excluded, but the name was changed to XTE~J1906+090 to match that commonly used in the literature.\\
\footnotetext[26]{It can be found in the recently published catalogue of high mass X-ray binaries in the Galaxy by \citet{2023A&A...671A.149F}.}\\
A handful of sources that are included in the recently published catalogue of high mass X-ray binaries by \citet{2023A&A...671A.149F} have been omitted from our catalogue. These sources have already been mentioned and are AX~J1700.1-4157, SAX~J1819.3-2525, AX~J1838.0-0655, and GRS~1758-258. In the case of GRS~1758-258, there is a possibility that the system is an intermediate-mass X-ray binary, as proposed by \citet{2016A&A...596A..46M}, and hence this source can be found in \citet{2023A&A...675A.199A} instead of this catalogue.

\section{Conclusion}
The Major changes with respect to \citet{2006A&A...455.1165L}, in addition to the problematic cases, are described here. 
\begin{itemize}
    \item  We homogenised the energy ranges of the reported X-ray fluxes. \citet{2006A&A...455.1165L} used an energy range of $2-10$\,keV for most sources, but in some cases the nominal energy range was different, and identifying these cases was not trivial. We now quote the soft and hard X-ray fluxes in well-defined energy ranges separately.
    \item \citet{2006A&A...455.1165L} only reported the maximum value of X-ray flux in units of Jy in most cases. In this work we report flux ranges in the two energy bands in the more commonly used cgs units, which are more accessible for X-ray sources.
    \item We increased the number of flags used to characterise source properties from 6 in \citet{2006A&A...455.1165L} to a total of 15 to better reflect the various features reported in the literature.
    \item While \citet{2006A&A...455.1165L} flags sources where a cyclotron line was detected, no information regarding the energy is available in most cases. We report both the observed energies and corresponding references. The number of objects where a line was detected also increased substantially due to the availability of new observations and publications.
    \item Effective stellar temperature and estimated optical luminosity, as well as the hydrogen column density are three completely new fields introduced in this version of the catalogue.
    \item Finally, we include up to six finding charts as part of the catalogue (from NIR images to hard X-ray bands) instead of  referencing the  existing finding charts.
\end{itemize}
Differences with respect to \citet{2023A&A...671A.149F} are as follows:
\begin{itemize}
    \item 
     \citet{2023A&A...671A.149F} utilised a flag to indicate the type of the optical companion. We used similar flags for the optical companion, but   introduced additional flags for the compact object (e.g. CRSF to indicate if a neutron star has a detected cyclotron line).
    \item 
    For sources with known cyclotron line energies we included both  fundamental and harmonic energies.
    \item 
    In addition to the cyclotron line energy, we also reported X-ray fluxes in different energy ranges for the compact object. This can be used to gauge the variability of the source and to make an estimation of the expected flux, although  this cannot replace an in-depth analysis of individual sources.  
    \item  
    For the optical companions, we reported magnitudes in the optical and in the NIR bands. Some additional information regarding the properties of optical counterparts can be found in \citet{2023A&A...671A.149F}. 

\end{itemize}
Together with the catalogues of  \cite{2023A&A...671A.149F} and \cite{2023A&A...675A.199A}, this catalogue provides a useful tool for further studies of XRBs. With ongoing surveys like eROSITA, the number of known XRBs is expected to increase substantially \citep{2014A&A...567A...7D}; therefore, we plan to keep the web version of our catalogue\footnote[27]{\url{http://astro.uni-tuebingen.de/~xrbcat/}}   up-to-date to the best of our ability.

\begin{acknowledgements}

This research has made use of the SIMBAD database and VizieR catalogue access tool operated at CDS, Strasbourg, France, and NASA’s Astrophysics Data System (ADS).
This research made use of hips2fits\footnotemark[15], a service provided by CDS.
AA thanks Deutsche Forschungsgemeinschaft (DFG) within the eROSTEP research unit under DFG project number 414059771 for support (DO 2307/2-1).
This work has made use of data from the European Space Agency (ESA) mission
{\it Gaia} (\url{https://www.cosmos.esa.int/gaia}), processed by the {\it Gaia}
Data Processing and Analysis Consortium (DPAC,
\url{https://www.cosmos.esa.int/web/gaia/dpac/consortium}). 
Funding for the DPAC
has been provided by national institutions, in particular the institutions
participating in the {\it Gaia} Multilateral Agreement. We acknowledge the public data from  \textit{XMM-Newton}, \textit{Chandra}, \textit{Swift} and \textit{INTEGRAL}.

\end{acknowledgements}
\bibliographystyle{aa}
\bibliography{bibtex.bib}

\newpage

\begin{appendix}

\onecolumn
\section{Catalogue format}

\begin{longtable}{p{0.2cm}p{3.2cm}p{2.4cm}p{11cm}}
\caption{\label{tab:cols} Definition of columns in the HMXB catalogue. In total the catalogue contains \textbf{66} columns.}\\

\hline 
№ & Column name & Unit & Description \\
\hline
\endfirsthead
\multicolumn{4}{c}%
{{\bfseries \tablename\ \thetable{} -- continued}} \\
\hline\hline
№ & Column name & Unit & Description \\ 
\hline
\endhead

\hline
\endlastfoot
    1 & `Name' &  &  Object name, which is the most common name in the literature
    \\

    2 & `RAdeg' &  deg  & Right Ascension in degrees (ICRS). 
     \\
  
    3 & `DEdeg' &  deg &  Declination in degrees (ICRS).
    \\
  
    4 &  `PosErr' &  arcsec & Positional error in arcseconds. 
    \\
  
    5 &  `Coord\_Ref' &  & Reference of catalogue or literature, which was used to extract the coordinates and uncertainties \\ 
    
    6 & `ID\_Flag' & & Robust identification of the optical counterpart: \\
    & & & 0 --- solidly identified optical counterpart in the literature \\
    & & & 1 --- tentative optical counterpart identification in the literature\\
    & & & 2 --- no identified optical counterpart\\

    7 &  `GLON' & deg & Galactic longitude in degrees. 
    \\

    8 & `GLAT' & deg &  Galactic latitude in degrees. 
    \\

    9 & `Xray\_Type' & &  List of the X-ray types assigned to the object. For more information see section above~\S~\ref{em: Flags}.
    \\

    10 &  `Porb' & day &  Orbital period of the binary system in days if determined.
         \\
    11 &  `Ppulse' & s &  Pulsation period (spin) of the binary (with NS) in seconds if determined.
    \\
    12 & `CRSF' & keV & Cyclotron line energies in keV
    \\

    13 &  `Alt\_Name' & & Second most used name in the literature.
    \\
    14 &  `SpType' & &  Spectral type of the optical counterpart.
    \\
  
    15 &  `Gaia\_DR3\_ID' & &  Source Name in the Gaia DR3 catalogue.
    \\
    16 &  `Gmag' & mag &  Optical magnitude in G-band according to a Gaia's catalogue. 
    \\     
    17 &  `e\_Gmag' & mag &  Corresponding  magnitude's error in G-band according to Gaia's catalogue. 
    \\     
    18 & `Vmag' & mag &  Optical magnitude in V-band according to SIMBAD database. 
    \\ 
    19 & `e\_Vmag' & mag &  Corresponding  magnitude's error in V-band according to SIMBAD database. 
    \\   
    20 & `Jmag' & mag &  IR magnitude in J-band according to 2MASS \citep{2003tmc..book.....C}.
    \\     
    21 & `e\_Jmag' & mag &  Corresponding  magnitude's error in J-band according to 2MASS \citep{2003tmc..book.....C}. 
    \\
    22 &  `Hmag' & mag&  IR magnitude in H-band according to 2MASS \citep{2003tmc..book.....C}.
    \\
    23 & `e\_Hmag' & mag &  Corresponding  magnitude's error in H-band according to 2MASS \citep{2003tmc..book.....C}.  
    \\
    24 & `Kmag' & mag &  IR magnitude in K-band according to 2MASS \citep{2003tmc..book.....C}.
    \\
    25 & `e\_Kmag' & mag &  Corresponding  magnitude's error in K-band according to 2MASS \citep{2003tmc..book.....C}.
    \\
    26 & `W1mag' & mag &  IR magnitude in W1 \textit{CatWISE}2020 band \citep{2021ApJS..253....8M}.  
    \\ 
    27 & `e\_W1mag' & mag &  Corresponding magnitude's error in W1 \textit{CatWISE}2020 band \citep{2021ApJS..253....8M}.
    \\ 
    28 & `W2mag' & mag &  IR magnitude in W2 \textit{CatWISE}2020 band \citep{2021ApJS..253....8M}. \\ 
    
    29 & `e\_W2mag' & mag &  Corresponding magnitude's error in W2 \textit{CatWISE}2020 band \citep{2021ApJS..253....8M}.
    \\   
    30 & `N\_H' & $10^{21}\ \mathrm{cm^{-2}}$ &  Neutral hydrogen column density, assuming a power-law spectrum \citep{2020ApJS..247...54E}. 
    \\
    31 & `XMM\_min\_flux' & $10^{-12}\ \mathrm{erg \ cm^{-2}\ s^{-1}}$ & Minimum X-ray flux for XMM-Newton (0.2-12.0 keV).
    \\
    32 & `XMM\_max\_flux' & $10^{-12}\ \mathrm{erg \ cm^{-2}\ s^{-1}}$ & Maximum X-ray flux for XMM-Newton (0.2-12.0 keV).
    \\
    33 & `Chandra\_min\_flux' & $10^{-12}\ \mathrm{erg \ cm^{-2}\ s^{-1}}$ & Minimum X-ray flux for Chandra ACIS (0.5-7.0 keV) or HRC (~0.1-10.0 keV).
    \\
    34 & `Chandra\_max\_flux' & $10^{-12}\ \mathrm{erg \ cm^{-2}\ s^{-1}}$ & Maximum X-ray flux for Chandra ACIS (0.5-7.0 keV) or HRC (~0.1-10.0 keV).
    \\
    35 & `Chandra\_Instrument' & & Chandra Instrument used for Chandra flux values.
    \\
    36 & `XRT\_min\_flux' & $10^{-12}\ \mathrm{erg \ cm^{-2}\ s^{-1}}$ & Minimum X-ray flux for Swift/XRT (0.3-10.0 keV).
    \\
    37 & `XRT\_max\_flux' & $10^{-12}\ \mathrm{erg \ cm^{-2}\ s^{-1}}$ & Maximum X-ray flux for Swift/XRT (0.3-10.0 keV).
    \\
    38 & `BAT\_min\_flux' & $10^{-12}\ \mathrm{erg \ cm^{-2}\ s^{-1}}$ & Minimum X-ray flux for Swift/BAT (14-195 keV).
    \\
    39 & `BAT\_max\_flux' & $10^{-12}\ \mathrm{erg \ cm^{-2}\ s^{-1}}$ & Maximum X-ray flux for Swift/BAT (14-195 keV).
    \\
    40 & `INTEGRAL\_min\_flux' & $10^{-12}\ \mathrm{erg \ cm^{-2}\ s^{-1}}$ & Minimum X-ray flux for INTEGRAL (17-60 keV).
    \\
    41 & `INTEGRAL\_max\_flux' & $10^{-12}\ \mathrm{erg \ cm^{-2}\ s^{-1}}$ & Maximum X-ray flux for INTEGRAL (17-60 keV).
    \\
    42 & `Mean\_Mass' & $\mathrm{M}_\odot $ & Mean mass of the compact object. \\
    43 & `Low\_Mass' & $\mathrm{M}_\odot$ & Lower limit of the compact object mass.\\
    44 & `High\_Mass' & $\mathrm{M}_\odot$ & Upper limit of the compact object mass.  \\
     
    45 & `Mean\_Dist' & pc &  Mean (between the two, minimum and maximum) estimate of the distance to the binary system \\ & & & according to {\citet{2023A&A...674A...1G,2021AJ....161..147B,2022A&A...658A..91A}} or taken from literature. In every literature case only one source of information was taken.
    \\
    46 & `Low\_Dist' & pc &  The lowest (minimum) estimate of the distance to the binary system according to {\citet{2023A&A...674A...1G,2021AJ....161..147B,2022A&A...658A..91A}} or taken from literature. In every literature  case only one source of information was taken.
    \\     
    47 & `High\_Dist' & pc & The highest (maximum) estimate of the distance to the binary system according to {\citet{2023A&A...674A...1G,2021AJ....161..147B,2022A&A...658A..91A}} or taken from literature. In every literature case only one source of information was taken. \\
    
    48 & `Teff' & K & Effective stellar temperature provided by {\citet{2023A&A...674A...1G,2022A&A...658A..91A}}.  \\
    49 & `Lopt' & $\mathrm{L}_\odot$ & Luminosity estimate of the optical companion provided by {\citet{2023A&A...674A...1G}}.\\
    50 & `IDS' & &  List of all identifiers, that we found in SIMBAD and in the catalogues and articles. \\
    51& `CRSF\_Ref' & & References to the articles, which providing the corresponding values for cyclotron line energies. \\
    52 & `Spin\_Ref'& & References to the articles, which providing the corresponding value for Pulsation Periods.  \\
    53 & `Orb\_Ref'& & References to the articles, which providing the corresponding value for Orbital parameters. \\
    54 & `Spectral\_Ref' & &References to the articles, which providing the spectral type of the optical star. \\
    55 & `Dist\_Ref' & & Reference to the article, which provides the distance estimation, if literature values were used.  \\
    56 & `Mass\_Ref' & & Reference to the article, which provides the mass estimation of the compact object. \\
    57 & `misc\_Ref' & & miscellaneous references regarding the object.\\
    58 & `comments' & & Comments regarding the object.  \\
    59 & `\_2MASS\_ID' & & Source name in the 2MASS All-Sky Catalogue of Point Sources\\
    60 & `CatWISE\_ID' & & Source name in the CatWISE2020 catalogue\\
    61 & `ROSAT\_ID' & & Source name in the Second ROSAT all-sky survey (2RXS) source catalogue\\
    62 & `XMM\_ID & & Source name in the XMM-Newton Serendipitous Source Catalogue\\
    63 & `Chandra\_ID' & & Source name in the Chandra Source Catalogue (CSC) Release 2.0\\
    64 & `XRT\_ID' & & Source name in the 2SXPS Swift X-ray telescope point source catalogue\\
    65 & `BAT\_ID' & & Source name in the Swift-BAT 105-Month All-Sky Hard X-Ray Survey catalogue\\
    66 & `INTEGRAL\_ID & & Source name in the INTEGRAL/IBIS 9-year Galactic Hard X-Ray Survey catalogue\\
\end{longtable}
\normalsize
\twocolumn
\end{appendix}

%
%

\end{document}